\documentclass[
reprint,
superscriptaddress,
amsmath,
amssymb,
aps,
nofootinbib
]{revtex4-2}

\usepackage{orcidlink}
\usepackage{amsmath}
\usepackage{hyperref}
\usepackage{mathtools}
\usepackage{color}
\usepackage{soul}
\sethlcolor{yellow}
\usepackage{tensor}
\usepackage{bm}

\usepackage{mathrsfs}

\def\sD{{\ensuremath{\mathcal{D}}}}
\def\sE{{\ensuremath{\mathcal{E}}}}

\begin{document}

\title{Upwind embedded boundary SBP operators: New high order numerical schemes for arbitrarily shaped domains with Cartesian grids}

\author{Conner Dailey \orcidlink{0000-0003-2488-3461}
}
\email[Corresponding Author: ]{conner.dailey@uni-jena.de}
\affiliation{Institute for Theoretical Physics, Friedrich-Schiller-Universit\"at, Jena, 07743, Germany}
\date{\today}

\begin{abstract}
    Embedded boundary summation by parts (SBP) methods define finite differencing based derivative operators with the added feature that the boundary need not coincide with a grid cell, allowing a boundary to be embedded on a regular Cartesian grid. This is achieved by the introduction of interpolation/extrapolation operators that match the accuracy of the boundary closure. These methods have been used to perform black hole excision simulations on a domain with a spherical boundary embedded in a regular Cartesian grid, demonstrating their usefulness for nonlinear problems. In this work, new operators are derived using this embedded boundary framework to increase the order of accuracy of the interior and boundary closure while minimizing the boundary error. Additionally, these novel operators improve the spectral properties on the grid by generalizing to an upwind scheme that has better dispersion relation preserving properties compared to traditional SBP schemes for wave equations. These operators are tested with the curvilinear scalar wave equation on a 3D multiblock grid with an excision sphere embedded in the center block to demonstrate the robustness and accuracy of these novel embedded operators.
\end{abstract}

\maketitle

\section{Introduction}\label{Sec:Intro}

A popular way to numerically solve hyperbolic partial differential equations (PDEs) are finite difference (FD) methods. In the context of solving initial boundary value problems (IBVPs), FD methods are attractive due to the locality of the derivative stencils. IBVP solvers, like many codes designed for modern CPUs and GPUs, are usually limited by available memory bandwidth, as compute capability has outpaced memory bandwidth improvements over the relatively recent history of computing hardware \cite{RooflineCPU,RooflineGPU,omlin2022highperformancexpustencilcomputations}. Thus a code implementation is often limited by how much memory must be loaded to calculate the derivative stencils, wider stencils being the main driver of increased computational cost. The locality of FD methods makes them relatively straight forward to scale to many devices, in contrast to spectral methods that require a comparatively large stencil size and are difficult to scale to many processors \cite{Pazos_2009}. An example of this straight forward scalability is the use of implicit global grids to distribute a FD domain over an arbitrary number of devices, which has been shown to achieve performance near theoretical hardware memory bandwidth limits and to scale to thousands of CPUs or GPUs \cite{omlin2022highperformancexpustencilcomputations,Omlin2024}.

Traditionally, FD methods are difficult to apply to IBVPs because the FD stencils must change in some consistent way so that only grid points in the problem domain are used when close to a boundary. This is part of the motivation for the summation by parts (SBP) framework, where FD operators can be defined consistently near the boundaries of the domain in such a way as to ensure the stability of linear hyperbolic problems using the energy method. The vast majority of SBP FD methods however require the last grid cell in the domain to coincide with the boundary. This is not much of a restriction in one dimension, but in 2D and 3D, care must be taken to ensure that the boundaries of the domain are properly captured by grid points. An example of this is the multiblock method, where a spherical outer boundary is achieved with rectangular arrays by applying a coordinate transform on several grid blocks (see for example \cite{SBP_2007} and \cite{Lehner_2005}). 

The multiblock method can also be used to achieve an inner excision boundary (see for example \cite{Schnetter_2006} and \cite{Pazos_2009}), but the clustering of grid cells near the inner boundary is inevitable, restricting the minimum stable time step that can be achieved. This method of obtaining boundary conforming grids cannot be applied as easily if one desires a \emph{dynamic} excision boundary, requiring the use of a sophisticated dual-frame evolution scheme \cite{Scheel_2006}. An example of the need for this is black hole excision in numerical relativity, where the problem solution in general can contain a singularity at the center of the black hole, so this region must be excised from the numerical domain so as not to spoil the numerical solution on the outside. Another compelling use case is that of worldtube excision, where an analytic perturbed solution on the inside of the excision boundary is coupled to the numerical domain via characteristic interface conditions \cite{Wittek_2023}. The applications for this type of excision boundary also include many other types of IBVPs, such as evolution equations subject to arbitrary boundary shapes with arbitrary boundary conditions (BCs), like those common in computational aeroacoustics \cite{DEUSE2020104449,computationalaeroacoustics}. One way to evolve IBVPs with arbitrarily shaped excision boundaries is with the embedded boundary SBP FD method of \cite{MATTSSON2017255}, which generalizes the traditional SBP FD framework to include operators that allow for boundaries that do not conform with the grid. In this way, a regular Cartesian grid can be used and an arbitrarily shaped boundary can be placed on that grid, with FD stencils that naturally adapt to the position of that boundary. The feasibility of this method for complex domain shapes and 2D linear hyperbolic problems was demonstrated in \cite{MATTSSON2017255}, and its application to the nonlinear Einstein equations in 3D was demonstrated in \cite{Dailey_2024}.

The operator derivations of \cite{MATTSSON2017255} were restricted to diagonal norm operators of interior order of accuracy ${p=2, 4, 6}$, which in this work are referred to as the $D_{2-1}$, $D_{4-2}$, and $D_{6-3}$ operators. As is well known in traditional SBP literature, the restriction to a diagonal norm requires a drop in accuracy of half the even interior order, hence the naming scheme. To improve upon these previously derived embedded boundary operators, this work derives and demonstrates several new diagonal norm upwind SBP operators that are generalized to embedded boundaries. This work generalizes the narrow stencil upwind operators of \cite{DiagonalNormUpwindSBP} and the dispersion relation preserving (DRP) upwind operators of \cite{DispersionRelationPreservingSBP}. Upwind schemes have been shown to improve the spectral properties on the grid when compared to traditional narrow stencil centered schemes, much more accurately approximating the dispersion relation of wave equations \cite{DualPairingWithDispersionPreserve,DispersionRelationPreservingSBP} while remaining explicit FD schemes. It has also been shown that this same DRP-type property can be realized in implicit (i.e. non-diagonal) FD schemes with the benefit of smaller interior stencil width \cite{DASZUTA2024112958}. These DRP-type FD schemes can ensure that the time of flight of waves on a numerical grid is much better approximated over long time scales. Upwind SBP schemes have also been shown to exhibit better than expected convergence properties than traditional SBP schemes which was shown experimentally in \cite{DiagonalNormUpwindSBP, mattsson2026optimalboundaryclosuresdiagonalnorm} and proven for specific BC application strategies in \cite{Jiang2024UpwindSBP}. Upwind schemes also allow for odd order accuracy in the interior, and the operators derived and explored in this work are of interior accuracy orders $p=2$ through $p=9$.

The use case for the embedded boundary method considered in this work is that of an inner excision boundary on a regular Cartesian grid, where the solution on the inside of this excised region is known either by some other type of numerical evolution scheme or via an analytical solution that is ``glued'' to the numerical domain. Thus the numerical tests are intended to demonstrate the feasibility of the novel operators when applied to one of two types of target problems. The first is black hole excision simulations, where the excision boundary lies within the apparent horizon, a completely inflowing surface, and thus the excision boundary does not require BCs if the characteristic speeds are causal. The other is worldtube excision, where an analytical solution on the interior of the excision region is coupled to the outer computational domain via characteristic interface conditions (see for example \cite{Wittek_2023}). The operators presented here can of course still be applied to more general domains and simulation types, but some careful implementation strategies due to geometric complications as outlined in \cite{MATTSSON2017255} may need to be applied, and are not tested here.

In Section~\ref{Sec:SBP}, traditional and embedded boundary SBP methods are reviewed  and in Section~\ref{Sec:Construction} the derivation of the novel operators is discussed. In Section~\ref{Sec:NumericalTests}, the operators are tested using the curvilinear wave equation on a multiblock grid with an excision sphere in the center. Section \ref{Sec:Conclusion} contains the conclusions and future work.

\section{Review of SBP Methods}\label{Sec:SBP}

The goal of SBP methods is to define consistent approximations of spatial derivatives while mimicking continuous integration by parts. With the help of SBP, problems that can be proven to contain a conserved continuous energy can also have a conserved numerical energy estimate on a discrete representation, allowing for a provably stable discrete evolution system via the energy method. An example of such a system of equations is the wave equation on a curvilinear domain with coordinates $x^i$, which can be written in first order form as
\begin{align}
\partial_t \psi&=-\Psi\,,\label{Eq.StaticSystem1}\\
\partial_t\psi_i + \partial_i\Psi&=0\,,\label{Eq.StaticSystem2}\\
\partial_t \Psi+\nabla_i\psi^i&= 0\,,\label{Eq.StaticSystem3}
\end{align}
where $\psi$ is the scalar wave solution, $\psi_i\equiv\partial_i\psi$, and $\Psi\equiv-\partial_t\psi$. Here, the divergence in curvilinear coordinates is written using Einstein summation convention as ${\nabla_i\psi^i=\partial_i(\sqrt{\gamma}\gamma^{ij}\psi_j)/\sqrt{\gamma}}$, where the spatial metric is $\gamma_{ij}$, its determinant is $\gamma$, and its matrix inverse is $\gamma^{ij}$. Given the continuous wave energy
\begin{align}
E\equiv \frac{1}{2}\int (\Psi^2+\psi^i\psi_i)\,dV\,,
\end{align}
where $dV$ is the volume element, one can show via continuous integration by parts (or equivalently the covariant divergence theorem)
\begin{align}
    \int u \nabla_iv^i\,dV+\int(\partial_i u)v^i\,dV=\oint u s_i v^i\,dS_B\,,\label{Eq:CovScalar}
\end{align}
for any scalar $u$ and vector $v^i$, that this energy satisfies the conservation law
\begin{align}
    \partial_tE = -\oint\Psi s_i\psi^i\,dS_B\,,
\end{align}
where $S_B$ are the boundary surfaces of the domain, $s^i$ are the outward pointing unit normal vectors to those surfaces, and $dS_B$ is the surface area element. This continuous problem is thus stable  for all time in the problem domain in the sense that this energy cannot grow with appropriate BCs, for example the reflecting Dirichlet condition $\Psi=0$.

The only requirement for this property to be mimicked in a discrete representation is that an analogue to continuous integration by parts exist in that discrete representation. For FD based SBP methods in 1D, if any $u$ and $v$ are sampled to a discrete grid, the SBP derivative operator $D$ and its associated norm $H$ are said to satisfy SBP when
\begin{align}
u^{\rm T}H Dv + (D u)^{\rm T}H v=u^{\rm T}Bv\,,\label{Eq.reg_SBP}
\end{align}
where $B$ is the boundary operator. Traditionally, ${B=\mathrm{diag}(-1,0,\dots,0,1)}$, which restricts the boundary flux to the positions of the last grid points in the domain. For now, $B$ remains unspecified as it will be generalized in the next subsection.

For the construction of centered SBP operators, it is convenient to define the antisymmetric matrix $Q$, which can be used to define the derivative operator
\begin{align}
D=H^{-1}\left(Q + \tfrac{1}{2}B\right)\,.
\end{align}
Inserting this definition into Eq.~(\ref{Eq.reg_SBP}) reveals that the pair $(Q,H)$ automatically satisfy SBP by construction, and this will be referred to as a $centered$ SBP scheme. Finding pairs $(Q,H)$ amounts only to enforcing the satisfaction of the polynomial accuracy constraints on $D$ for an appropriate order of boundary stencil accuracy. Forcing $Q$ to have the property $Q=-Q^T$ necessarily forces the interior stencil to be a centered FD stencil, however this need not be the case in general.

Instead one might require two different matrices $Q^\pm$ to satisfy $Q^+=-(Q^-)^T$ and define two separate derivative operators
\begin{align}
D^\pm=H^{-1}\left(Q^\pm + \tfrac{1}{2}B\right)\,,
\end{align}
which satisfy a different but equally valid SBP rule by construction
\begin{align}
u^{\rm T}H D^\pm v + (D^\mp u)^{\rm T}H v=u^{\rm T}Bv\,.\label{Eq.dual_SBP}
\end{align}
Thus, the triplet $(Q^\pm,H)$ also satisfies SBP, and this is known as a \emph{dual pairing} SBP scheme \cite{DualPairSBP}. This triplet has some useful properties, namely that  
\begin{align}
D_1 &\equiv \tfrac{1}{2}(D^++D^-)\,,\\ H^{-1}S &= \tfrac{1}{2}(D^+-D^-)\,,
\end{align}
where $D_1$ is always a centered SBP operator (although not necessarily a narrow stencil one) that satisfies Eq.~(\ref{Eq.reg_SBP}). If the matrix $S$ is additionally negative semi-definite, the triplet $(Q^\pm,H)$ is known as an \emph{upwind} SBP scheme \cite{DiagonalNormUpwindSBP}. This added property produces a centered SBP and numerical dissipation pair $(D_1, H^{-1}S)$, which is very useful as a separate numerical dissipation operator need not be constructed after the fact. When these operators are used, which can be thought of as equivalent to Steger–Warming flux splitting \cite{DiagonalNormUpwindSBP}, the scheme is referred to here as a \emph{centered upwind} scheme. The $D^\pm$ operators can also be applied asymmetrically in a first order hyperbolic system, for example using the $D^+$ operator for the divergence in Eq.~(\ref{Eq.StaticSystem3}) and using the $D^-$ operator for the gradient in Eq.~(\ref{Eq.StaticSystem2}) which is referred to here as an \emph{asymmetric upwind} scheme. This type of scheme greatly improves the spectral properties on the grid compared to centered schemes, see for example  \cite{DualPairingWithDispersionPreserve} and \cite{DispersionRelationPreservingSBP}, where it is shown that the dispersion relation of wave equations can be preserved much more accurately compared to centered schemes. For this scheme type, the phase velocity of the highest resolvable frequency on the grid is nonzero unlike all centered schemes, so numerical dissipation is not strictly necessary in the interior as numerical noise propagates instead of accumulating over time. However, the absence of numerical dissipation can still lead to instabilities near boundaries especially for nonlinear problems. Another feature of the asymmetric scheme is the absence (or minimization) of spurious wave modes, a phenomenon in which high frequency modes have negative group velocity and thus wave packets propagate in the incorrect direction. It was shown in \cite{DualPairingWithDispersionPreserve} that some operators do not support these spurious modes in the asymmetric scheme, and all of the operators in that work lessen their existence compared to centered schemes, while \cite{DispersionRelationPreservingSBP} eliminates them almost entirely.

Traditionally, the interior stencil is taken to be of minimal width for a given order of interior accuracy. This does not necessarily need to be the case, as the parameters in an interior FD stencil can be optimized for properties other than accuracy. For example, the interior stencils of \cite{DispersionRelationPreservingSBP} are wider than is minimally required for the accuracy order, but the free parameters are then optimized for minimum dispersion error, called a dispersion relation preserving (DRP) scheme. These schemes far outperform in dispersion error tolerance compared to the minimal width upwind schemes when applied asymmetrically, which comes at the cost of a larger interior stencil width. 

\begin{figure}[tbp]
\includegraphics[width=0.95\linewidth]{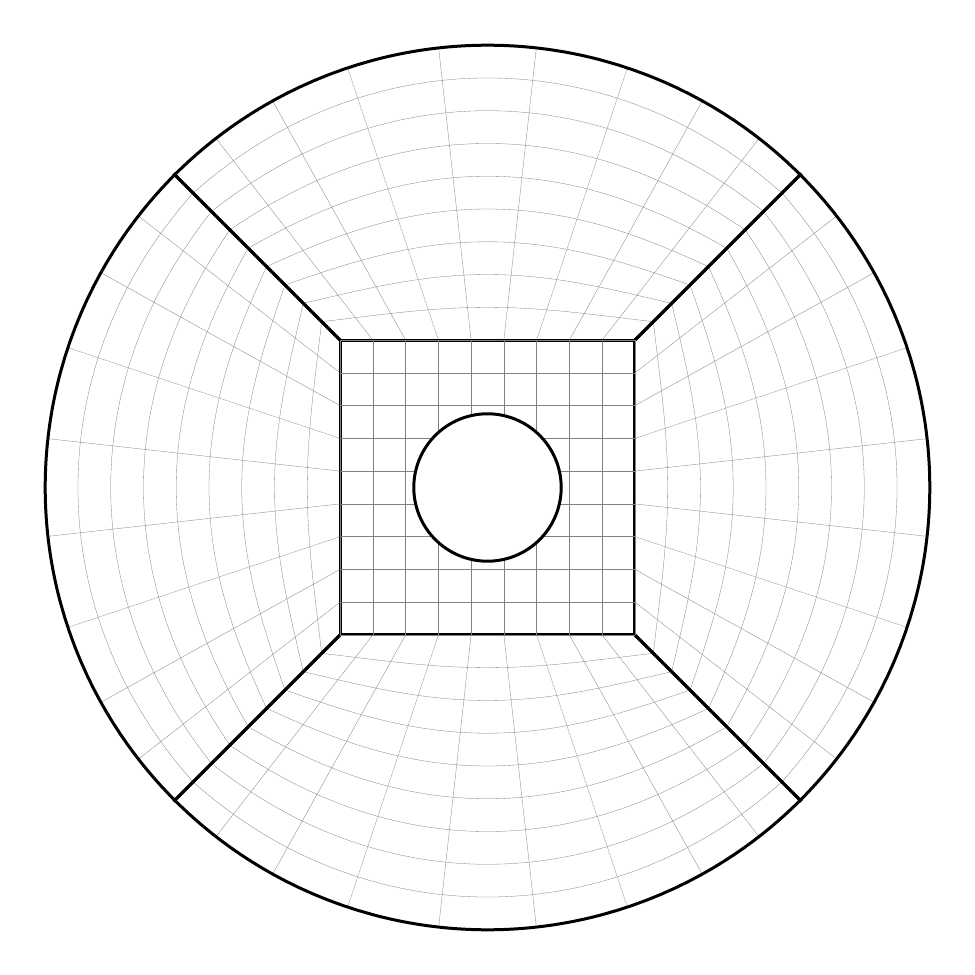}
\caption{Equatorial slice of the test domain grid for a resolution of $10^3$ grid points per block. It consists of seven blocks in a cubed-sphere topology with a central sphere excised in the central block via the embedded boundary method.}\label{Fig:MeshBlocks}
\end{figure}

\subsection{Embedded Boundary SBP Methods}

The embedded boundary SBP FD method of \cite{MATTSSON2017255} is defined in 1D relative to an equidistantly spaced grid with arbitrarily placed boundary points. The coordinates of the $n$ computational grid points are given by 
\begin{align}
x_k = x_1 + (k-1)h\,,
\end{align}
for integer grid index $1\leq k\leq n$, initial coordinate grid value $x_1$, and coordinate grid spacing $h$. The positions of the left and right boundaries are given by  
\begin{align}
x_l = x_1-\alpha_l h\,,\quad x_r = x_n+\alpha_r h\,,
\end{align}
respectively, where $\alpha_{l,r}$ measure the signed distance of the boundary \emph{away} from the first/last grid point in units of the grid spacing, i.e. a positive value indicates the boundary is \emph{outside} of the computational domain, while a negative value indicates the boundary lies between the first/last two active grid points.

To obtain values on the boundary points, the interpolation/extrapolation operators are introduced. Following the notation of \cite{DiagonalNormUpwindSBP}, if $X_q$ is a vector containing grid samples of the function $x^q/q!\,$, then the $p^{\rm th}$ order accurate interpolation/extrapolation vectors are defined as those that satisfy
\begin{align}
e_l^TX_q = x_l^q/q!\,,\quad e_r^TX_q = x_r^q/q!\,,
\end{align}
for all integers $q\in [1,p]$ at the left and right boundaries respectively. Then the traditional SBP property Eq.~(\ref{Eq.reg_SBP}) and the upwind property Eq.~(\ref{Eq.dual_SBP}) are generalized by defining the boundary operator
\begin{align}
B=-e_l e^T_l+e_r e^T_r\,.
\end{align}
If the desired boundary accuracy order is $b$, then $e_{l,r}$ are chosen to be $b^{\rm th}$ order accurate vectors.

With $B$ specified, 1D operators can be derived to arbitrary order of interior accuracy in principle. When generalizing to multi-dimensions, the technique is to define the problem line by line, effectively treating the domain as a collection of 1D problems. As was discussed in \cite{MATTSSON2017255}, the geometry of a multidimensional problem can lead to several complications which are briefly reviewed.

\emph{Trapped Points} When the boundary is nearly parallel to a coordinate direction, the computational domain might contain trapped points that belong to one line, but not to others due to the allowed range in $\alpha_{l,r}$. This issue can be resolved by extrapolating derivatives in whatever direction they cannot be applied normally.

\emph{Slender Geometries} If two boundaries are nearly touching each other, it is possible that a point belongs to several stencil lines at once. This issue can be resolved by introducing two copies of that point and evolving each separately. 

\emph{Narrow Domains} If the outer boundary has sharp features, such as the corner of a triangle, or if two boundaries are sufficiently near each other, grid points may not have enough neighbors along the active lines in the computational domain to apply appropriately sized FD stencils for the given interior order of accuracy. In this case, the order of accuracy must either be locally decreased or the derivative must be extrapolated. 

In the numerical tests demonstrated in Section~\ref{Sec:NumericalTests}, none of these geometry complications are present, likely improving the stability of the overall scheme. The slender geometry and the narrow domain problems are not present due to the design of the numerical domain being constrained such that they do not appear. The trapped point problem will in general be a complication for an inner excision boundary for an IBVP, and the method of derivative extrapolation used in \cite{MATTSSON2017255} will need to be applied. However, the target use case for these methods is situations where the solution on the interior of the excision boundary is known, either by an analytically known solution or by some other numerical evolution method. For this use case, points on the inside have known time evolution separate from the main computational domain, which can be used to avoid the trapped point problem.

\section{Embedded Operators}\label{Sec:Construction}

\subsection{Construction Details}

Here, the construction sequence of the various operators defined in this paper is elaborated. The diagonal norm upwind scheme is characterized by a diagonal ansatz for $H$ and a $2b\times2b$ boundary block on the upper left and bottom right sections of $Q^+$, with $Q^-=-(Q^+)^T$. The interior stencil is taken to be that of an upwind biased $p^{\rm th}$ order accurate FD stencil, such as those given in \cite{DiagonalNormUpwindSBP} or \cite{DispersionRelationPreservingSBP}. In the case of $p=2$, the minimal ansatz takes the form
\begin{align}
Q^+ &= \left(\begin{array}{ccccc}
 q_{11} & q_{12}  & -\tfrac{1}{2} & 0&\cdots\\
  q_{21}   & q_{22} & 2&  -\tfrac{1}{2} &   \\
  0 & 0 & -\frac{3}{2} & 2 & -\tfrac{1}{2}\\
  \vdots   &     & &&\ddots
\end{array}\right)\,,\nonumber\\ H&=h
\left(\begin{array}{cccc}
 h_{11} & 0    & 0 & \cdots \\
 0  & h_{22} &  0      & \\
  0   &   0& 1      & \\
  \vdots &  &  &\ddots
\end{array}\right)\,.
\end{align}
As SBP is satisfied by construction, the only constraints on the yet unspecified parameters are the $2(b+1)$ accuracy conditions:
\begin{align}
(Q^++\tfrac{1}{2}B)X_q - HX_{q-1}&=0\,,\\
(Q^-+\tfrac{1}{2}B)X_q - HX_{q-1}&=0\,,
\end{align}
for all integers $q\in [1,b]$. 
However, the accuracy conditions alone are not always sufficient to define a unique operator, and thus there may be free parameters in the solution. To set these values, the  boundary error can be minimized. For boundary accuracy order $b$, the $q^{\rm th}$ order boundary error vector for each of the $D^\pm$ operators is defined as
\begin{align}
\sE^\pm_q=(Q^\pm+\tfrac{1}{2}B)X_{q} - HX_{q-1}\,,
\end{align}
for $q>b$ and the $L_2$ error norm is $||\sE_q^\pm||^2=h(\sE_q^\pm)^T \sE_q^\pm$. The functions $||\sE_q^\pm||^2$ depend quadratically on the free parameters, and can be minimized with respect to them. The  error norms are used to form the total boundary error function:
\begin{align}\label{Eq:MinObjective}
f=\sum^{2b-1}_{q=b+1}\left(||\sE_q^+||^2+||\sE_q^-||^2\right)\,, 
\end{align}
which can then be minimized with respect to the free parameters available.

It should be noted that the range of $\alpha_{l,r}$ is not necessarily restricted a priori. Originally in \cite{MATTSSON2017255} it was defined as ${-1/2\leq\alpha_{l,r}<1/2}$, but this need not be the case in general, and the highest order operator norms in this work are not even positive definite in this range. As long as a full range is defined, i.e. $\alpha_{\rm max}-\alpha_{\rm min}=1$, then the embedded scheme is complete in the sense that any boundary position can be properly represented on the grid, and $|\alpha_{l,r}|<1$ as the grid can always be adjusted so that the coordinates of the first/last grid points are within a distance $h$ from the boundary positions. One may be tempted to define for example $0\leq\alpha_{l,r}<1$, as this strategy is completely extrapolated, and thus immune to the trapped point problem for a convex inner boundary. The immediate issue with this when operators are derived is that the norm is typically not positive definite in this range for operators of interior order $p>3$. One also discovers that the boundary error typically increases sharply for $\alpha_{l,r}>0$, so extrapolation should be avoided if possible to reduce the overall simulation error. Another important consideration is the maximum spectral radius over the chosen $\alpha_{l,r}$ interval, as this directly affects the maximum stable time step of an explicit Runge-Kutta evolution. 

The construction sequence is thus:
\begin{enumerate}
    \item Construct a parameterized ansatz for $(Q^+,H)$ given an upwind interior stencil
    \item Construct $B$ with $b^{\rm th}$ order accurate $e_{l,r}$ vectors
    \item Find a solution of $(Q^+,H)$ that satisfies the $2(b+1)$ accuracy conditions
    \item Minimize the total error function $f$ with respect to the remaining free parameters
    \item Ensure that the matrix $S$ is negative semi-definite
    \item Select a range $[\alpha_{\rm min},\alpha_{\rm max})$ for $\alpha_{\rm max}-\alpha_{\rm min}=1$ for which the norm is positive definite and the maximum spectral radius is minimized
\end{enumerate}

In this work, solutions were found for interior orders $p=2$ through $p=9$ using the interior stencils of \cite{DiagonalNormUpwindSBP} and these operators are labeled as $D^\pm_{p-b}$ for interior order $p$ and boundary order $b$. Solutions for the DRP interior stencils of \cite{DispersionRelationPreservingSBP} are labeled $\sD^\pm_{p-b}$.

\begin{table}[tbp]
    \centering
    \begin{tabular}{|c|c|c|c|c|}\hline\hline
                        Operator &  $[\alpha_{\rm min},\alpha_{\rm max})$ & $\rho^{(1)}_{\rm max}$ & $\rho^{(2)}_{\rm max}$ & $\rho^{(3)}_{\rm max}$ \\ \hline
        $D^\pm_{2-1}$   &   $[\,0,1)$     & 2.70  & 3.86 & 5.55  \\
        $D^\pm_{3-1}$   &   $[-1/5,4/5)$     & 1.21 & 1.45 & 1.92  \\
        $D^\pm_{4-2}$   &   $[-1/2,1/2)$  & 5.04 & 4.63 & 4.92  \\ 
        $D^\pm_{5-2}$   &   $[-1/2,1/2)$  & 1.48 & 1.58 & 1.59  \\ 
        $D^\pm_{6-3}$   &   $[-3/5,2/5)$  & 1.89 & 2.18 & 2.80  \\
        $D^\pm_{7-3}$   &   $[-3/5,2/5)$  & 1.66 & 1.69 & 1.75  \\
        $D^\pm_{8-4}$   &   $[-2/3,1/3)$  & 2.02 & 2.40 & 2.54  \\
        $D^\pm_{9-4}$   &   $[-2/3,1/3)$  & 2.04 & 2.05 & 2.08 \\\hline
        $\sD^\pm_{4-2}$   &   $[-1/2,1/2)$  & 2.19 & 2.63 & 3.62  \\ 
        $\sD^\pm_{5-2}$   &   $[-1/2,1/2)$  & 1.75 & 2.69 & 3.63  \\ 
        $\sD^\pm_{6-3}$   &   $[-3/5,2/5)$  & 2.05 & 2.77 & 3.80  \\
        $\sD^\pm_{7-3}$   &   $[-3/5,2/5)$  & 1.93 & 2.80 & 3.80  \\\hline
    \end{tabular}
    \caption{Properties of all derived operators. Each operator comes with designed boundary position range in ${\alpha_{l,r}\in[\alpha_{\rm min},\alpha_{\rm max})}$. For that whole range, the maximum occurring spectral radius of each of the three discussed scheme types is listed as $\rho^{(i)}_{\rm max}$, where $(1)$ is the centered upwind scheme, $(2)$ is the asymmetric scheme, and $(3)$ is the asymmetric dissipative scheme.}
    \label{Tab:OpProps}
\end{table}

\begin{figure}[tbp]
\includegraphics[width=\linewidth]{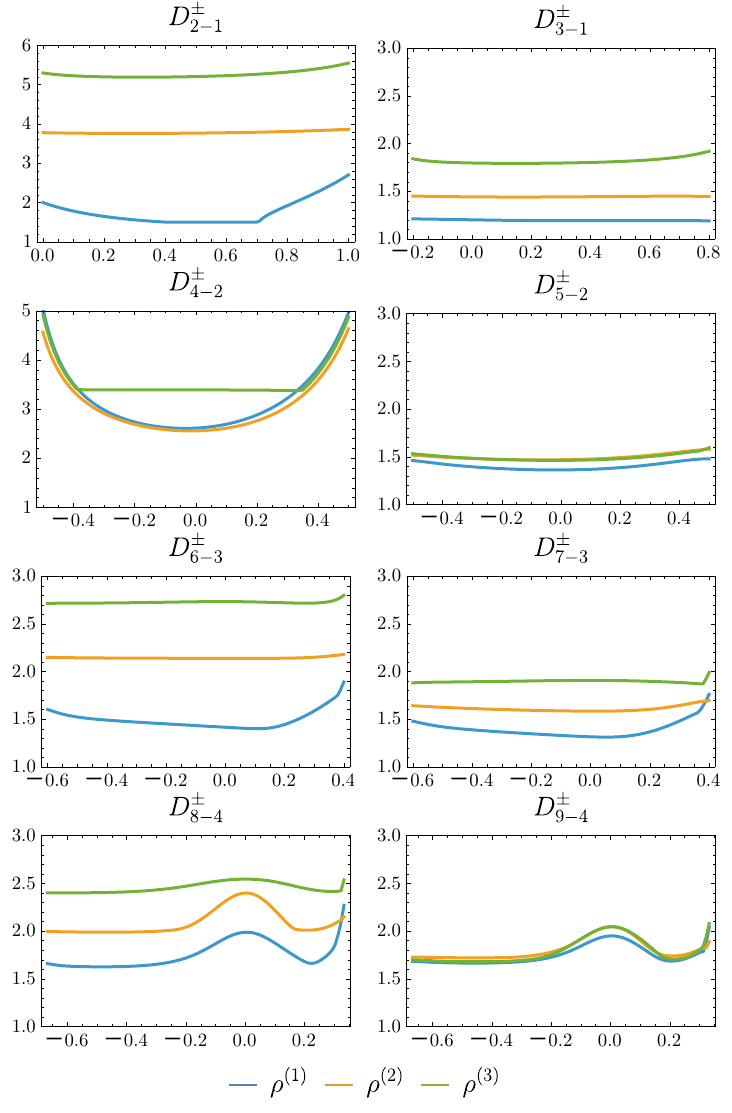}
\caption{Spectral radii $\rho^{(i)}$ for the $D^\pm_{p-b}$ series of operators as a function of the boundary position parameter $\alpha$ over the defined range in Table~\ref{Tab:OpProps} when applied via the three discussed scheme types with characteristic BCs. Here, $\rho^{(1)}$ corresponds to the centered upwind scheme, $\rho^{(2)}$ to the asymmetric upwind scheme, and $\rho^{(3)}$ for the asymmetric dissipative scheme. Even order operators tend to have bad values that improve as the order increases, while the odd order operators tend to have better values that cluster closer for the different schemes. Note the vertical axis limits for the $D^\pm_{2-1}$ and $D^\pm_{4-2}$ operators are higher than the rest.}\label{Fig:SectralRadii}
\end{figure}

\begin{figure}[tbp]
\includegraphics[width=\linewidth]{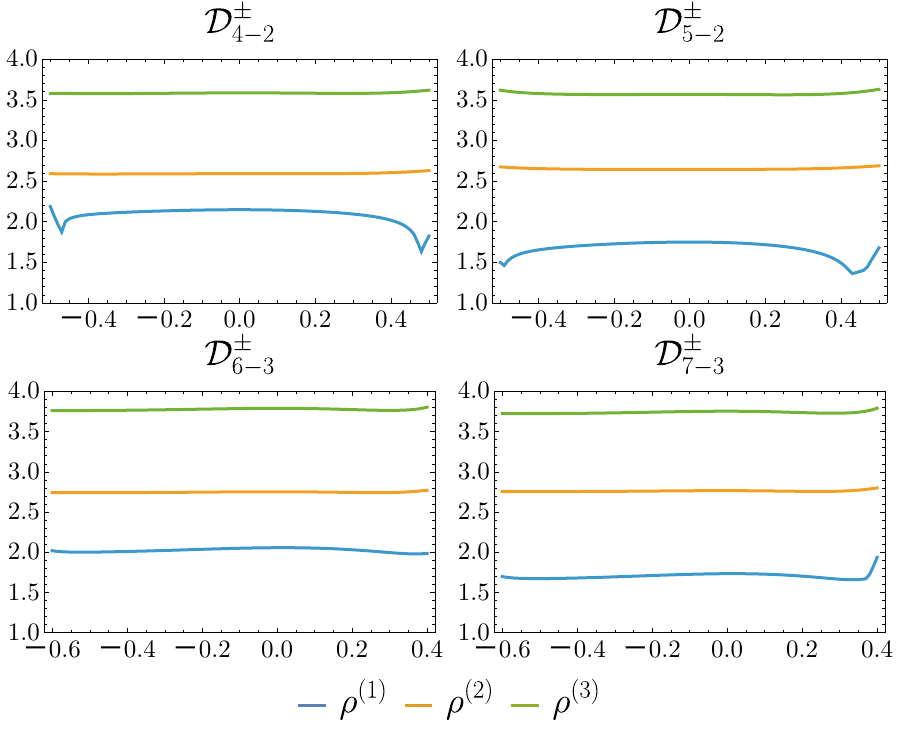}
\caption{Spectral radii $\rho^{(i)}$ for the $\sD^\pm_{p-b}$ series of operators as a function of the boundary position parameter $\alpha$ over the defined range in Table~\ref{Tab:OpProps} when applied via the three discussed scheme types with characteristic BCs. Here, $\rho^{(1)}$ corresponds to the centered upwind scheme, $\rho^{(2)}$ to the asymmetric upwind scheme, and $\rho^{(3)}$ for the asymmetric dissipative scheme. In contrast to the $D^\pm_{p-b}$ series of operators, the spectral radii vary little between operators.}\label{Fig:SectralRadiiDRP}
\end{figure}

\subsubsection{Minimal Width Upwind Operators}

The $D^\pm_{2-1}$ and $D^\pm_{3-1}$ operators are unique in the minimal boundary closure, but surprisingly the norm for $D^\pm_{2-1}$ is only positive definite in the range ${-0.3\lesssim\alpha_{l,r}\lesssim1.6}$, and the spectral radius is best in the purely extrapolating range ${0\leq\alpha_{l,r}<1}$. The norm of the $D^\pm_{3-1}$ operator is positive definite in the range ${-0.6\lesssim\alpha_{l,r}\lesssim1.4}$, and the range ${-1/5\leq\alpha_{l,r}<4/5}$ minimizes the spectral radius. Both schemes can be made free from the trapped point problem for a convex inner boundary in principle by selecting ${0\leq\alpha_{l,r}<1}$. The matrix $S$ is naturally negative semi-definite in both cases. The calculation of the spectral radii is detailed in Section~\ref{Sec:OpProps}.

The $D^\pm_{4-2}$ and $D^\pm_{5-2}$ operators have one free parameter each, the norm is positive definite in the range ${-0.6\lesssim\alpha_{l,r}\lesssim0.8}$, and the spectral radius is best in the range ${-1/2\leq\alpha_{l,r}<1/2}$ for both cases. The free parameter can be used to minimize the total error function, and this results in a unique minimum and $S$ is naturally negative semi-definite for this minimum.

The $D^\pm_{6-3}$ and $D^\pm_{7-3}$ operators have four free parameters each, and the norm is positive definite in the range ${-0.7\lesssim\alpha_{l,r}\lesssim0.6}$. Minimizing the error function with respect to these 4 free parameters results in a unique minimum and a negative semi-definite $S$ in both cases. The resulting spectral radii can be optimized by selecting the range ${-3/5\leq\alpha_{l,r}<2/5}$ for both. 

The $D^\pm_{8-4}$ and $D^\pm_{9-4}$ operators have nine free parameters each, and the norm is positive definite in the range ${-0.7\lesssim\alpha_{l,r}\lesssim0.4}$. Minimizing the error function with respect to these 9 free parameters results in a unique minimum and a negative semi-definite $S$ in both cases. Selecting ${-2/3\leq\alpha_{l,r}<1/3}$ seems to minimize sharp increases in the spectral radii for both. 

It should be noted that in all cases, while the resulting solutions for $H$ and $Q^+$ are polynomials of order $2b$ in $\alpha_{l,r}$, the matrix $S$ is always independent of $\alpha_{l,r}$, which is very convenient. It should also be noted that for operators of interior order $p>5$, taking $\alpha_{l,r}=0$ does not exactly reproduce to the solutions of \cite{DiagonalNormUpwindSBP}, although the free parameters are numerically within a few percent. This is likely due to slightly different error minimization methodologies. 

\subsubsection{Dispersion Relation Preserving Operators}

The $\sD^\pm_{4-2}$ and $\sD^\pm_{5-2}$ operators have one free parameter each, the norm is positive definite in the range ${-0.6\lesssim\alpha_{l,r}\lesssim0.8}$, and the spectral radius is best in the range ${-1/2\leq\alpha_{l,r}<1/2}$ for both cases. The free parameter can be used to minimize the total error function, and this results in a unique minimum and $S$ is naturally negative semi-definite for this minimum. These design properties are nearly identical to the $D^\pm_{4-2}$ and $D^\pm_{5-2}$ operators.

The $\sD^\pm_{6-3}$ and $\sD^\pm_{7-3}$ operators have four free parameters each, and the norm is positive definite in the range ${-0.7\lesssim\alpha_{l,r}\lesssim0.6}$. Minimizing the error function with respect to these 4 free parameters results in a unique minimum and a negative semi-definite $S$ in both cases. The resulting spectral radii can be optimized by selecting the range ${-3/5\leq\alpha_{l,r}<2/5}$ for both, again resulting in design properties are nearly identical to the $D^\pm_{6-3}$ and $D^\pm_{7-3}$ operators.

It should be noted again that taking $\alpha_{l,r}=0$ does not exactly reproduce to the solutions of \cite{DispersionRelationPreservingSBP}, again likely due to slightly different error minimization methodologies. 

The properties of all derived operators are given in Table~\ref{Tab:OpProps}, including the selected range in $\alpha_{l,r}$ as well as the spectral radii for different scheme types elaborated in the next section.  This work includes supplementary material in plain text for the reader's convenience that lists the rational polynomial coefficients in $\alpha_{l,r}$ for all derived operators as well as a Mathematica notebook that can be used to re-derive and symbolically manipulate the solutions.

\subsection{Numerical Scheme Types}\label{Sec:OpProps}

\begin{figure}[tbp]
\includegraphics[width=0.95\linewidth]{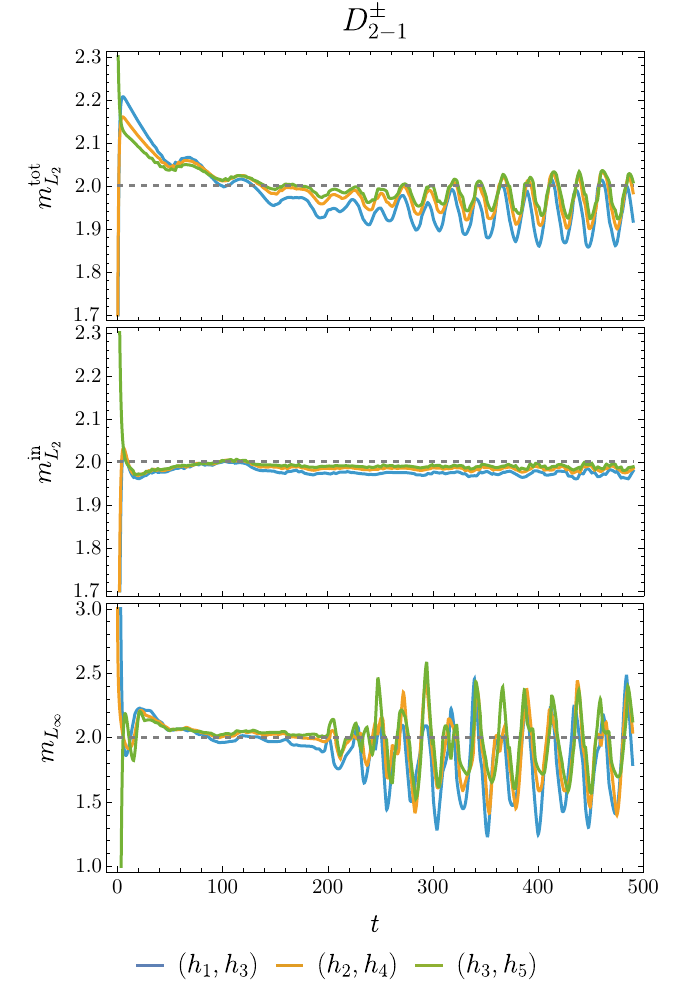}
\caption{Convergence results for the $D^\pm_{2-1}$ operators. Dashed lines indicate expected convergence orders, and the numerical results cluster closely to these values.}\label{Fig:D_2-1}
\end{figure}

\begin{figure}[tbp]
\includegraphics[width=0.95\linewidth]{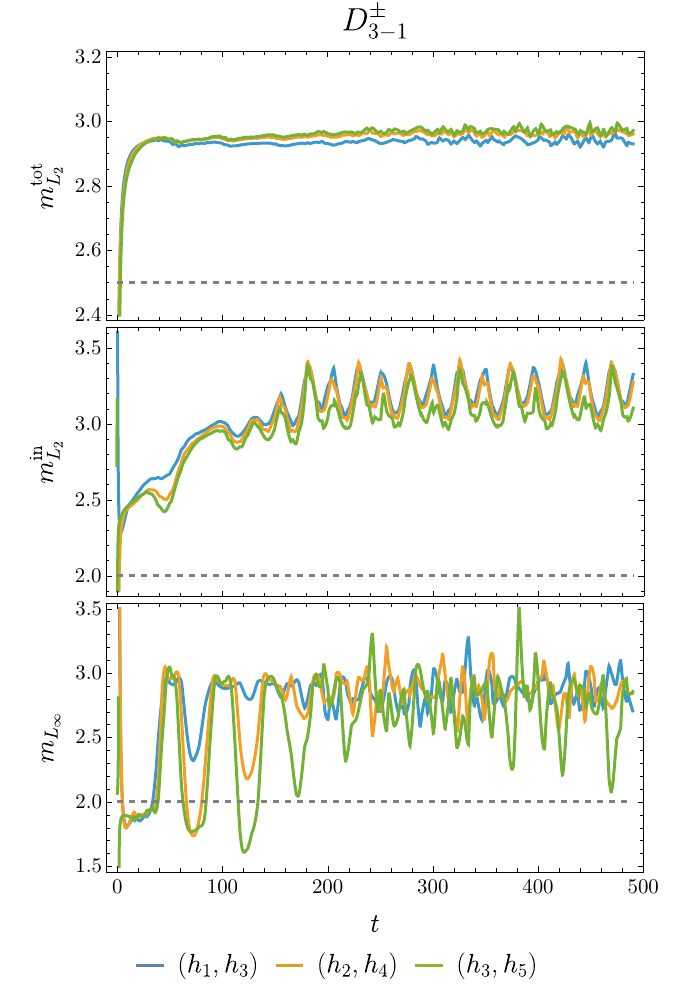}
\caption{Convergence results for the $D^\pm_{3-1}$ operators. Dashed lines indicate expected convergence orders, and these operators overperform these expectations substantially.}\label{Fig:D_3-1}
\end{figure}

\begin{figure}[tbp]
\includegraphics[width=0.95\linewidth]{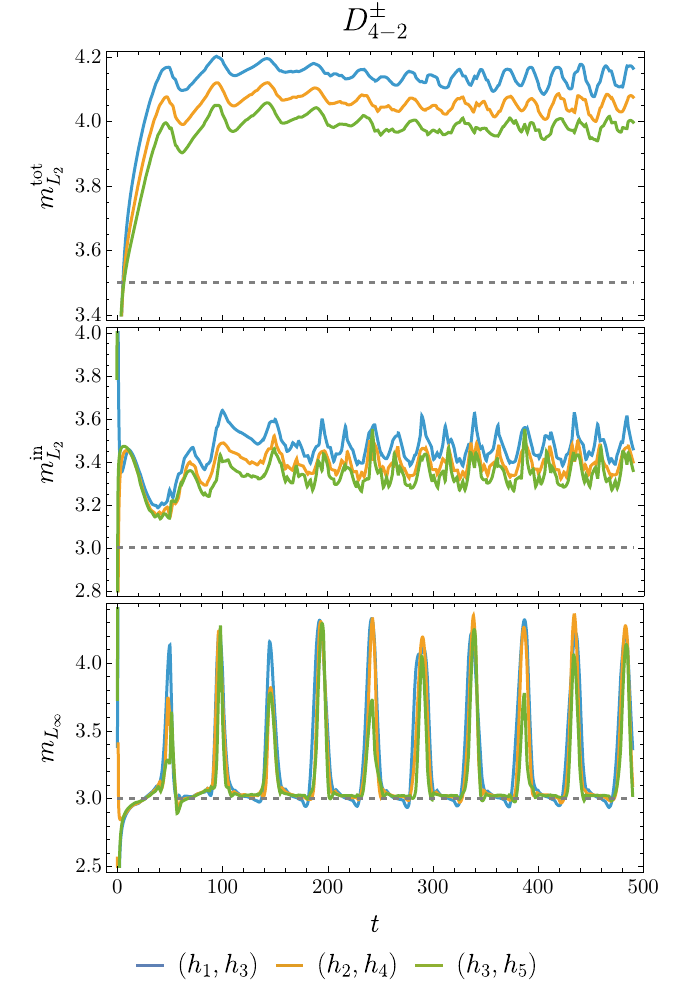}
\caption{Convergence results for the $D^\pm_{4-2}$ operators. Dashed lines indicate expected convergence orders, and these operators overperform these expectations at least pre-asymptotically.}\label{Fig:D_4-2}
\end{figure}

\begin{figure}[tbp]
\includegraphics[width=0.95\linewidth]{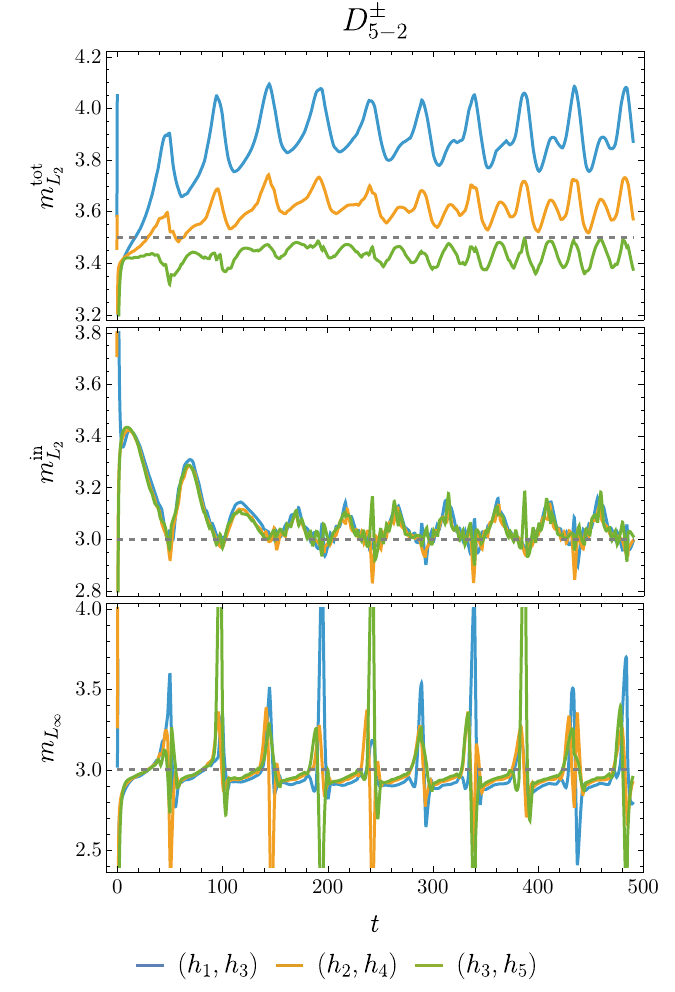}
\caption{Convergence results for the $D^\pm_{5-2}$ operators. Dashed lines indicate expected convergence orders, and these operators overperform initially, but quickly settle near the expectations.}\label{Fig:D_5-2}
\end{figure}

\begin{figure}[tbp]
\includegraphics[width=0.95\linewidth]{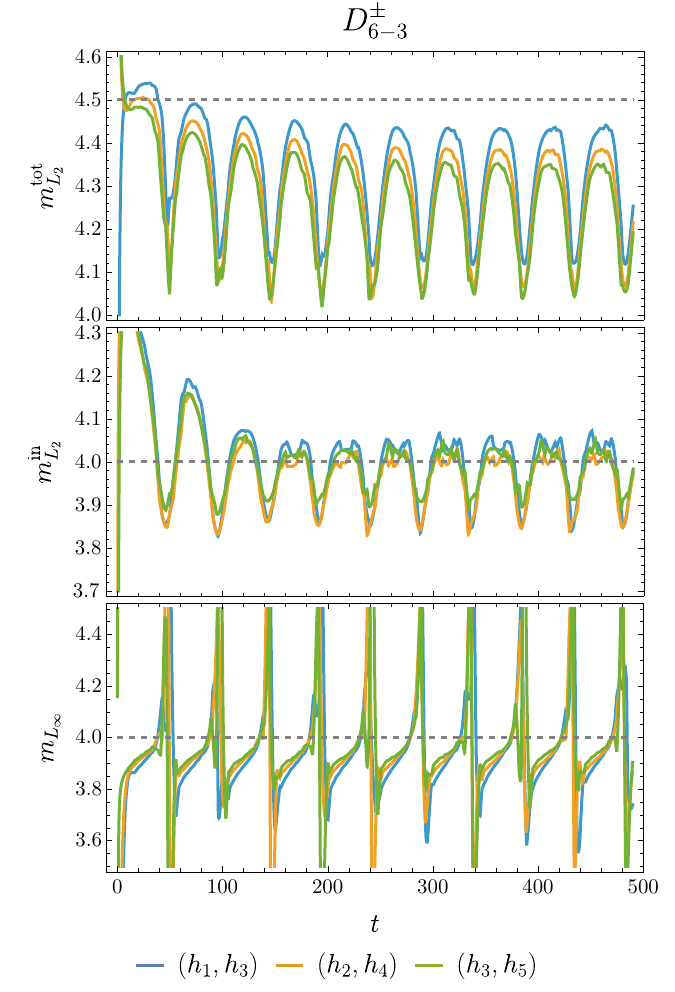}
\caption{Convergence results for the $D^\pm_{6-3}$ operators. Dashed lines indicate expected convergence orders, with numerical results aligning with these expectations.}\label{Fig:D_6-3}
\end{figure}

\begin{figure}[tbp]
\includegraphics[width=0.95\linewidth]{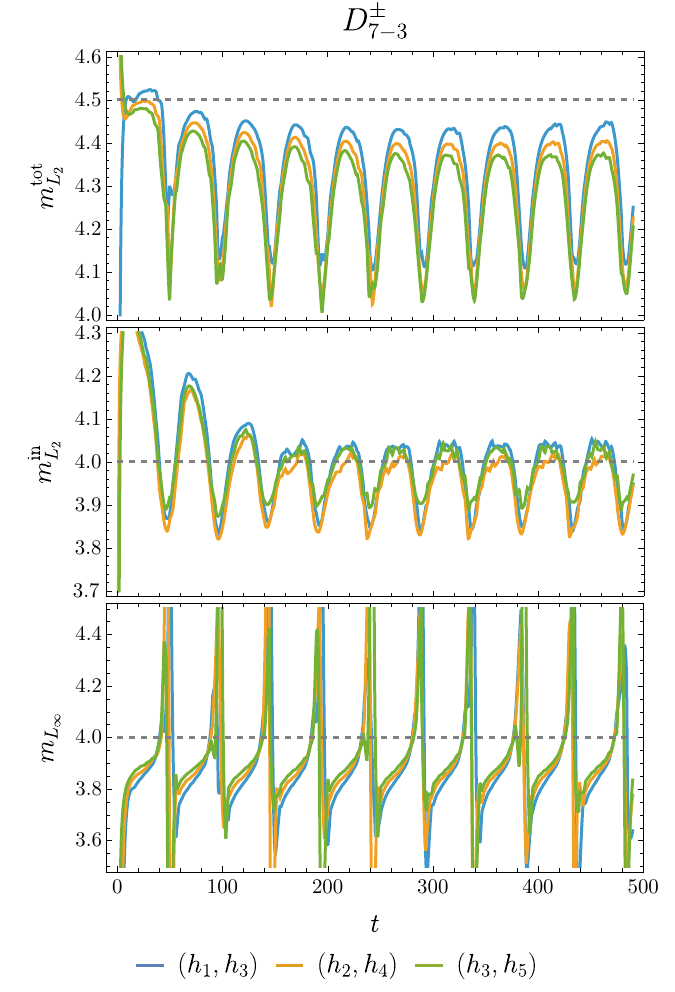}
\caption{Convergence results for the $D^\pm_{7-3}$ operators. Dashed lines indicate expected convergence orders, with numerical results aligning with these expectations.}\label{Fig:D_7-3}
\end{figure}

\begin{figure}[tbp]
\includegraphics[width=0.95\linewidth]{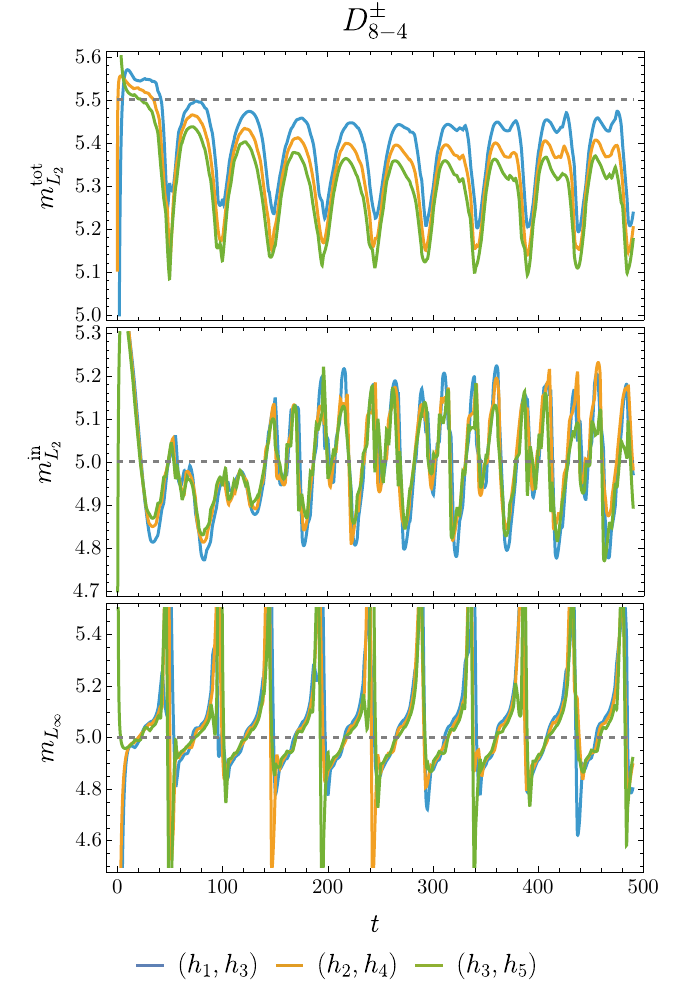}
\caption{Convergence results for the $D^\pm_{8-4}$ operators.  Dashed lines indicate expected convergence orders, with numerical results aligning with these expectations.}\label{Fig:D_8-4}
\end{figure}

\begin{figure}[tbp]
\includegraphics[width=0.95\linewidth]{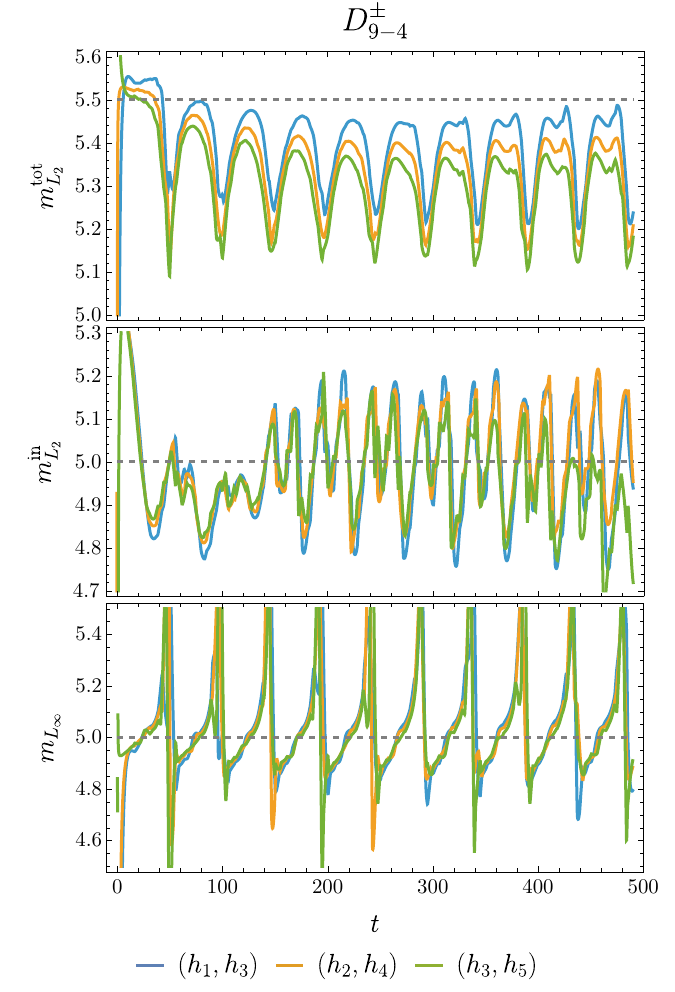}
\caption{Convergence results for the $D^\pm_{9-4}$ operators. Dashed lines indicate expected convergence orders, with numerical results aligning with these expectations.}\label{Fig:D_9-4}
\end{figure}

\begin{figure}[tbp]
\includegraphics[width=0.95\linewidth]{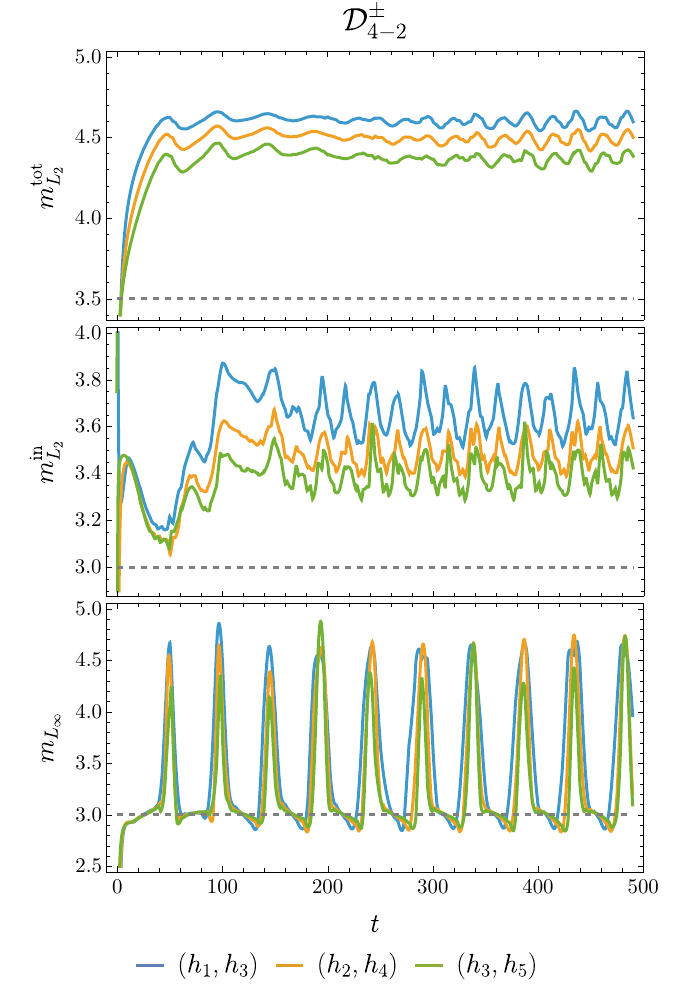}
\caption{Convergence results for the $\sD^\pm_{4-2}$ operators. Dashed lines indicate expected convergence orders, and the numerical results overperform these expectations at least pre-asymptotically}\label{Fig:D_4-2_disp}
\end{figure}

\begin{figure}[tbp]
\includegraphics[width=0.95\linewidth]{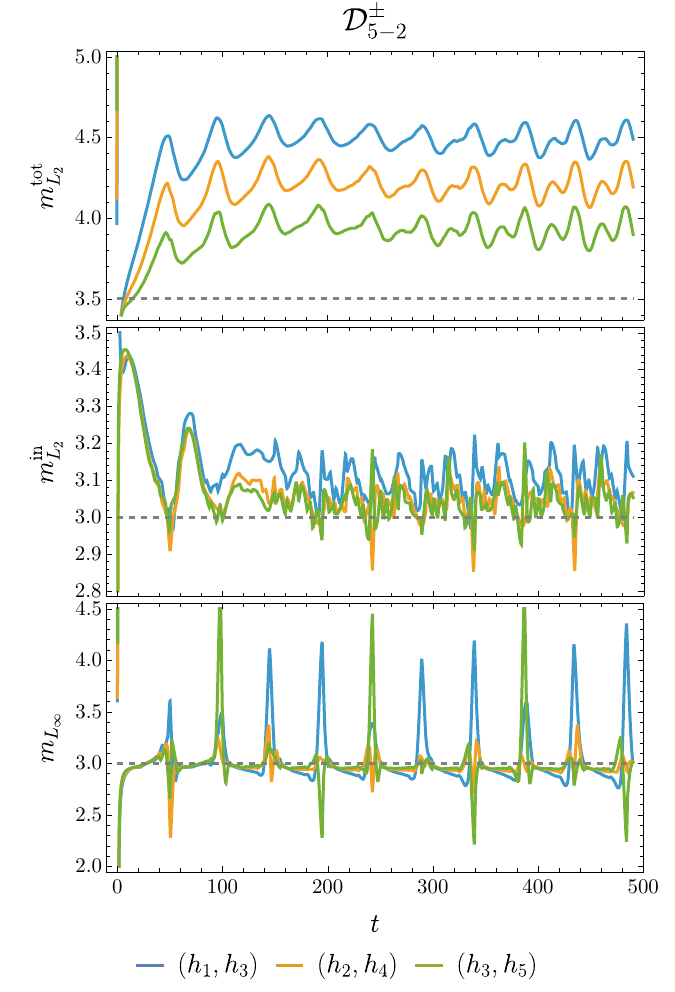}
\caption{Convergence results for the $\sD^\pm_{5-2}$ operators. Dashed lines indicate expected convergence orders, and the numerical results overperform these expectations at least pre-asymptotically. }\label{Fig:D_5-2_disp}
\end{figure}

\begin{figure}[tbp]
\includegraphics[width=0.95\linewidth]{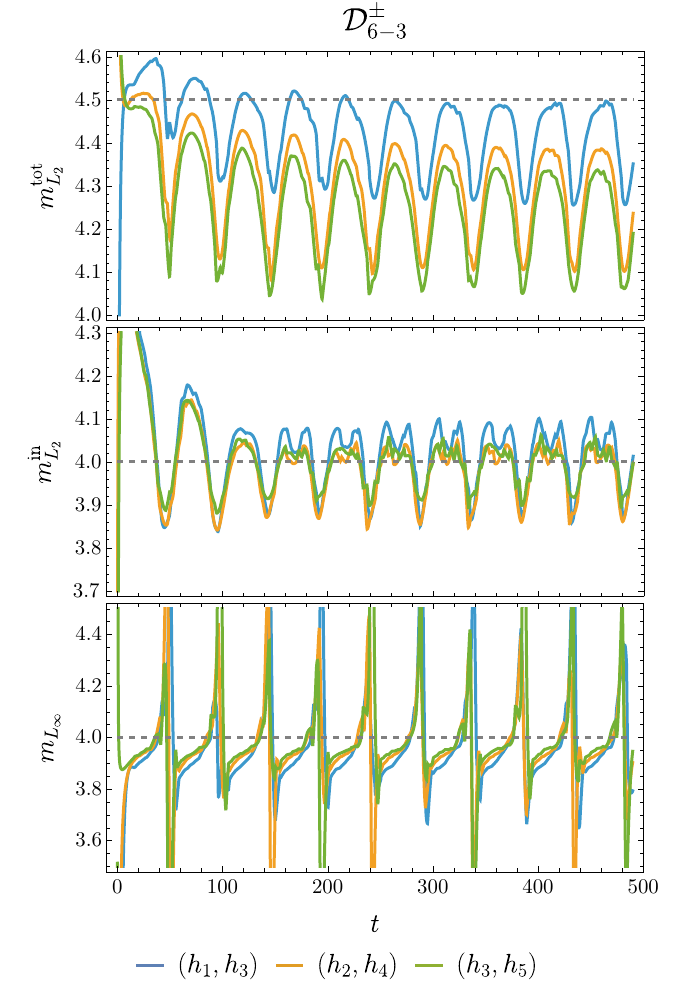}
\caption{Convergence results for the $\sD^\pm_{6-3}$ operators. Dashed lines indicate expected convergence orders, and the numerical results cluster close to these expectations.}\label{Fig:D_6-3_disp}
\end{figure}

\begin{figure}[tbp]
\includegraphics[width=0.95\linewidth]{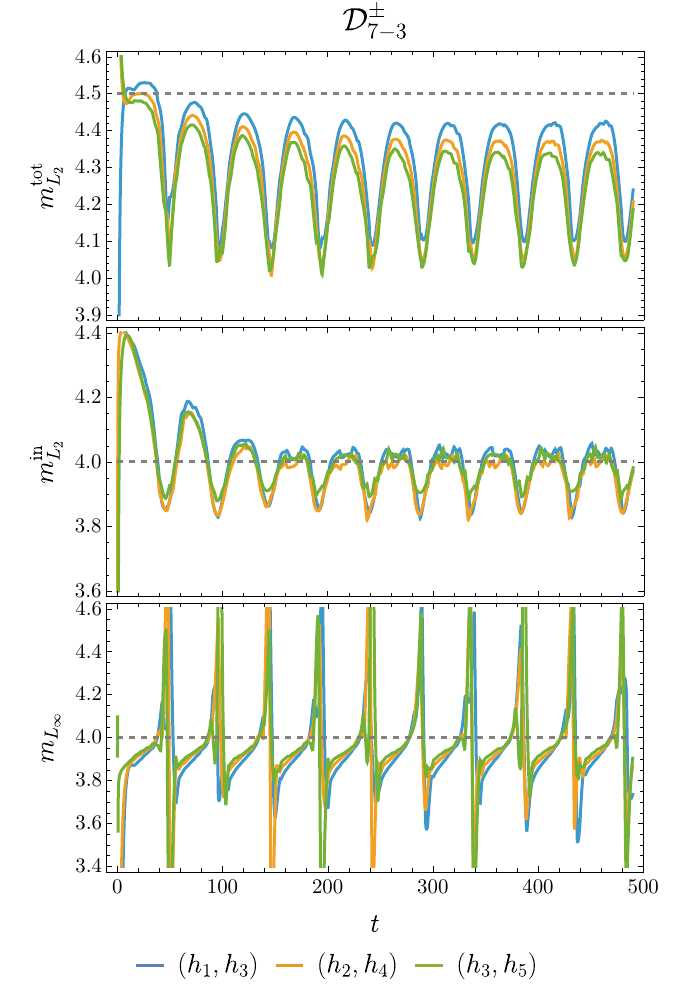}
\caption{Convergence results for the $\sD^\pm_{7-3}$ operators. Dashed lines indicate expected convergence orders, and the numerical results cluster close to these expectations.}\label{Fig:D_7-3_disp}
\end{figure}

To explore the properties of these new operators for hyperbolic problems in first order form, the following simplest hyperbolic system is considered:
\begin{align}\label{Eq:Simple}
\partial_tU = A \partial_x U 
\end{align}
where $U=(u,v)$ is the state vector sampled on a 1D grid with $n$ points. The principle matrix and its characteristic decomposition are given by 
\begin{align}
A=\bigg(
\begin{array}{cc}
    0 & 1 \\
    1 & 0
\end{array}
\bigg),\, A_+=\frac{1}{2}\bigg(
\begin{array}{cc}
    1 & 1 \\
    1 & 1
\end{array}
\bigg),\, A_-=\frac{1}{2}\bigg(
\begin{array}{cc}
   -1 & 1 \\
    1 & -1
\end{array}
\bigg)\,,
\end{align}
and for future convenience, the following matrices are defined:
\begin{align}
I_2=\bigg(
\begin{array}{cc}
    1 & 0 \\
    0 & 1
\end{array}
\bigg),\, A_1=\bigg(
\begin{array}{cc}
    0 & 1 \\
    0 & 0
\end{array}
\bigg),\, A_2=\bigg(
\begin{array}{cc}
    0 & 0 \\
    1 & 0
\end{array}
\bigg)\,.
\end{align}
It is also useful to define the Kronecker product
\begin{align}
A\otimes D\equiv\bigg(
\begin{array}{cc}
    A_{11}D & A_{12}D  \\
    A_{21}D  & A_{22}D 
\end{array}
\bigg)\,.
\end{align}
In \cite{DiagonalNormUpwindSBP}, the application of the upwind operators is taken as a Steger-Warming flux splitting technique with Dirichlet BCs. In this work, only characteristic BCs are considered, and under homogeneous characteristic BCs the semi-discretization of Eq.~(\ref{Eq:Simple}) reads
\begin{align}
\partial_tU &= A_+\otimes D^+U + A_-\otimes D^-U\nonumber\\ &+ A_+\otimes H^{-1} e_l U_l- A_-\otimes H^{-1} e_r U_r\,.
\end{align}
where $U_{l,r}\equiv e^T_{l,r}U$. It was shown in \cite{DiagonalNormUpwindSBP} that this scheme is equivalent to the centered scheme with numerical dissipation:
\begin{align}
\partial_tU &= A \otimes D_1 U + I_2\otimes (H^{-1}S)U\nonumber\\ &+ A_+\otimes H^{-1} e_l U_l- A_-\otimes H^{-1} e_r U_r\,.
\end{align}
Thus, this scheme will be referred to as a \emph{centered upwind} scheme. We can also apply the $D^\pm$ operators asymmetrically as was done in \cite{DispersionRelationPreservingSBP, DualPairingWithDispersionPreserve}. This scheme reads
\begin{align}
\partial_tU &= A_1\otimes D^- U + A_2\otimes D^+U\nonumber\\ &+ A_+\otimes H^{-1} e_l U_l- A_-\otimes H^{-1} e_r U_r\,.
\end{align}
Switching $A_1$ with $A_2$ here makes no practical difference to the scheme. This scheme is referred to as an \emph{asymmetric upwind} scheme and it can be shown to approximate the dispersion relation of wave equations much more accurately, where the phase velocity of the highest resolvable frequency on the grid is non-zero \cite{DispersionRelationPreservingSBP, DualPairingWithDispersionPreserve}. This is an attractive feature, as all centered schemes have zero phase velocity at this mode, so numerical noise does not propagate on the interior grid in this case. However, numerical dissipation is still an important feature to have to stabilize boundaries in the multidimensional case and in general for nonlinear problems. It can be easily added back by modifying the asymmetric scheme as 
\begin{align}
\partial_tU &= A_1\otimes D^- U + A_2\otimes D^+U+I_2\otimes (H^{-1}S)U\nonumber\\ &+ A_+\otimes H^{-1} e_l U_l- A_-\otimes H^{-1} e_r U_r\,.
\end{align}
This scheme is referred to as an \emph{asymmetric dissapative} scheme. This type of scheme is attractive from the perspective of boundary stability, as if unstable modes are produced near a boundary, they propagate and damp over time, preventing instabilities from accumulating near boundaries by two separate mechanisms. One of the advantages to the multiblock approach is that one can select a scheme type for each block individually in principle.

In 1D all three of these schemes are provably stable with the energy method thanks to the upwind SBP property. This cannot be proven however in the case of 2D or 3D in the presence of embedded boundaries, and the hope is that with appropriate numerical dissipation, the scheme can be stabilized. In principle, one can increase the strength of the numerical dissipation for both the centered upwind scheme and the asymmetric dissipative scheme by introducing a parameter greater than one in front of the dissipation term, but that was not found to be necessary in this work. More challenging domain geometries, such as those with sharp corners, may require this to obtain a stable scheme near the boundary.

\section{Numerical Tests}\label{Sec:NumericalTests}

\subsection{Domain design}

The embedded boundary framework allows for an arbitrarily shaped domain in 2D and 3D with a Cartesian grid, with specific details outlined in \cite{MATTSSON2017255}. For example, a finite spherical shell can be defined using one 3D Cartesian grid array, with an outer spherical boundary and an inner spherical excision of desired radii. The embedded boundary operators then naturally adjust to the position of the boundaries wherever the grid lines may intersect them, but with some complications outlined at the end of Section~\ref{Sec:SBP}. However, there are reasons to prefer a multiblock setup to achieve a given outer boundary shape rather than using an embedded boundary method. Taking the finite spherical shell as an example, an outer boundary necessarily will introduce the narrow domain problem where the outer boundary is nearly parallel with the Cartesian coordinate directions, forcing the use of smaller stencils that will reduce the local order of accuracy, while the multiblock method does not suffer from this problem. Also, a naive implementation would use a full Cartesian array and the largest sphere in the corresponding coordinate space, wasting significant computer memory for grid points outside the evolution domain. Another important design feature is the ability to push the outer boundary farther away from the center by the use a radial coordinate transform, allowing much larger domains to be represented with the same memory footprint. For these reasons, tests are only performed with an inner excision surface. 

Furthermore, the numerical tests are intended to demonstrate the properties of these new operators when the solution on the inner excision surface is known, either by some other numerical evolution scheme or by an analytical solution coupled to the computational domain. This follows the  intended use case for these methods to excise regions in a numerical relativity simulation. For these types of simulations, the inner solution can be injected into the computational domain via incoming characteristic modes wherever BCs are required, while points inside the boundary have known time evolution. To gain the spectral benefits of asymmetric schemes for wave systems, as well as introduce numerical dissipation in anticipation of these methods being applied to nonlinear problems, the numerical tests performed here exclusively use the asymmetric dissipative scheme type. 

The design of the test IBVP follows \cite{SBP_2007} closely, so as to make a direct comparison to the same domain without an embedded boundary. The domain is a cubed sphere multiblock topology with inner block side length $l$ and outer radius $R=3l/2$. On its own, this domain is provably stable with diagonal norm SBP operators and SAT interface conditions. An inner spherical excision of radius $l/4$ at the center of the domain is introduced as an inner boundary, visualized in Figure~\ref{Fig:MeshBlocks}. The reference solution to the wave equation is identical to that of \cite{SBP_2007}, a plane wave traveling in the direction $k^i$ with frequency $\omega^2 = k^i k_i$:
\begin{align}
    \psi(t,x^i) = A\cos[2\pi(k^i x_i-\omega t)]\,.
\end{align}
The initial conditions are evaluated at $t=0$ and the auxiliary variables $\Psi$ and $\psi_i$ are initialized via their definitions. Compared to the domain setup of \cite{SBP_2007}, lengths and times are scaled by a factor of 50, such that the inner block has side length $l=100$. For all test runs, the initial conditions are set with $A=1$ and ${k^x=k^y=k^z=0.6/50}$. The frequency is chosen  higher than \cite{SBP_2007}, as the $D^\pm_{9-4}$ operator must have at least 16 points for a complete operator, and the minimum resolution for the domain setup is set by this requirement, leading to a higher minimum resolution compared to that work. The increased frequency assures that roughly the same number of points per wavelength is being tested.

The inner and outer BCs are injected via incoming characteristics relative to this same reference solution. While the characteristics at the multiblock interfaces are defined relative to the surface normals at those interfaces, the embedded boundary is defined \emph{line by line}, so characteristics are defined relative to the Cartesian grid line that intersects the embedded boundary surface. This is exactly the 3D tensor product version of the numerical schemes defined in Section~\ref{Sec:OpProps}. Points sufficiently inside the boundary are not active, and thus not evolved, while points just inside the boundary within a distance $|\alpha_{\rm min}|h$  along grid lines are active. These active points inside the boundary are evolved via the exact time derivatives obtained by differentiating the reference solution. 

\begin{figure}[tbp]
\includegraphics[width=\linewidth]{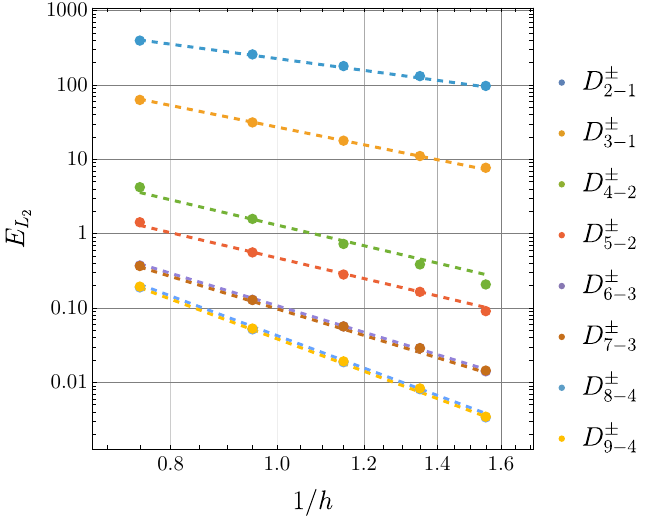}
\caption{Comparison of the convergence of the $D^\pm_{p-b}$ series of operators via the total domain $L_2$ error. For boundary order $b$, dashed lines are proportional to the expected convergence rate of $b+1.5$ for each operator, some of which appear to over-perform this expected rate.}\label{Fig:total_convergence}
\end{figure}

\begin{figure}[tbp]
\includegraphics[width=\linewidth]{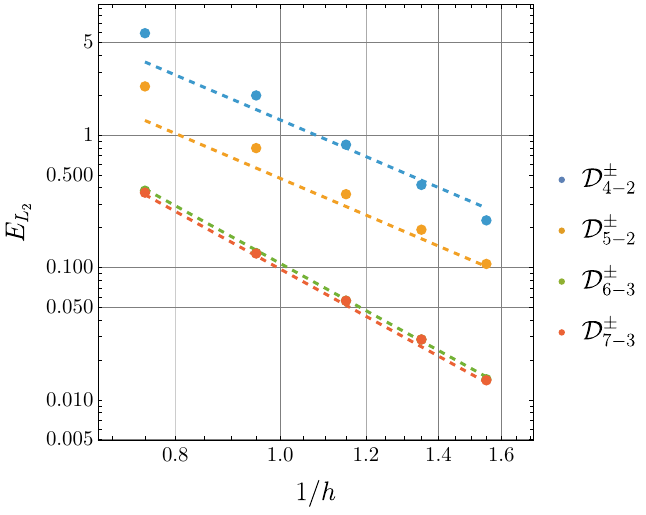}
\caption{Comparison of the convergence of the $\sD^\pm_{p-b}$ series of operators via the total domain $L_2$ error. For boundary order $b$, dashed lines are proportional to the expected convergence rate of $b+1.5$ for each operator, and the height of the lines are identical to those of the respective $D^\pm_{p-b}$ series of operators in Figure~\ref{Fig:total_convergence} to facilitate direct comparison.}\label{Fig:total_disp_convergence}
\end{figure}

\subsection{Test Evolutions}

To test the convergence properties of the operators, the convergence order is defined as
\begin{align}
    m = \frac{\log\left(E_{h_1}/E_{h_2}\right)}{\log(h_1/h_2)}\,,
\end{align}
where $E_h$ is the solution error calculated with various types of norm over only active grid points for a given grid spacing $h_i$. If the error is calculated with the $L_2$ norm on all blocks, the convergence order is labeled $m^{\rm tot}_{L_2}$. If the error is calculated with the $L_2$ norm only in the central block, it is labeled $m^{\rm in}_{L_2}$. If instead the $L_\infty$ norm is used over all blocks, it is labeled $m_{L_\infty}$. These different metrics help give a sense of the convergence to be expected in different regions of the domain. The $m^{\rm tot}_{L_2}$ gives the total convergence and showcases the expected $b+1.5$ asymptotic convergence order that occurs with upwind operators \cite{Jiang2024UpwindSBP}. The $m^{\rm in}_{L_2}$ gives the convergence to be expected when the domain is heavily dominated by boundary error, which tends be at the usual traditional SBP asymptotic convergence order of  $b+1$. Finally, $m_{L_\infty}$ shows the worst case convergence over the whole domain, also with an expected asymptotic value of $b+1$, although it frequently jumps as the worst case grid cell convergence can switch discontinuously over time. Five test resolutions are chosen ($h_{1}-h_{5}$) that have $n=76,96,116,136,156$ grid points per block dimension respectively. Time stepping is performed using the $5^{\rm th}$ order accurate explicit Runge-Kutta scheme $\rm{ERK}(20,5)$ of \cite{ParsaniOptimizedRK} to ensure that the highest order operators are not limited by the accuracy of the time stepping method, with a time step size of $h_t=0.8h$.

\subsubsection{Minimal Width Operators}

\emph{The $D^\pm_{2-1}$ Operators } The $p=2$ case is the lowest order option possible and it requires an interior stencil width of 5. In a departure from the general trend of traditional SBP operators, the $p=2$ case has the worst spectral radius properties by far, requiring comparatively small maximum time steps compared to all other operators for all three schemes. The convergence results in Figure~\ref{Fig:D_2-1} show second order convergence regardless of the error monitoring method.

\emph{The $D^\pm_{3-1}$ Operators } The $p=3$ case shows an improvement of spectral radius of roughly a factor of 3 for all schemes, making it much more favorable to the $D^\pm_{2-1}$ operators for time stepping efficiency. Surprisingly, the convergence results in Figure~\ref{Fig:D_3-1} indicate third order convergence regardless of the error monitoring method, higher than expected. As the $p=3$ case has the same stencil width as the $p=2$ case, it seems to be the better choice for a scheme of interior stencil width 5.

\emph{The $D^\pm_{4-2}$ Operators } The first scheme to require an interior stencil width of 7, the $p=4$ case has very bad spectral radius properties, some schemes are worse than the $p=2$ case. Its performance makes up for this with higher than expected convergence rates in the $L_2$ based errors as shown in Figure~\ref{Fig:D_4-2}. It is observed that $m^{\rm tot}_{L_2}\sim 4$ and $m^{\rm tot}_{L_2}\sim 3.5$, although these rates decay slightly over resolution increase, indicating they are pre-asymptotic. This unexpectedly good convergence may be connected to the fact that this operator has the lowest dispersion error for an asymmetric scheme as can be seen in \cite{DualPairingWithDispersionPreserve}. This trend is seen also in the following section on the DRP operators.

\emph{The $D^\pm_{5-2}$ Operators } The $p=5$ case shows a significant improvement of spectral radius over the $p=4$ case, roughly a factor of 3 again for all three scheme types, in fact it has the lowest max value of all operators for the asymmetric dissipative scheme. Its convergence rates are closer to the expected values as shown in Figure~\ref{Fig:D_5-2}. The $p=4$ case actually outperforms the convergence rate of the $p=5$ case at least at the resolutions tested here, although the initial error for $p=5$ is far lower, resulting in lower total $L_2$ error overall as seen in Figure~\ref{Fig:total_convergence}.

\emph{The $D^\pm_{6-3}$ Operators } The first scheme to require an interior stencil width of 9, the $p=6$ case has worse spectral radii than the $p=5$ case.  Figure~\ref{Fig:D_6-3} shows convergence values that lie very close to the expected values of $m^{\rm tot}_{L_2}\sim4.5$ and $m^{\rm in}_{L_2}\sim m_{L_\infty}\sim4$.

\emph{The $D^\pm_{7-3}$ Operators } The $p=7$ case shows an improvement in the spectral radii for only some scheme types over the $p=6$ case.  Figure~\ref{Fig:D_7-3} shows practically identical convergence rates to the $p=6$ case.

\emph{The $D^\pm_{8-4}$ Operators } The first scheme to require an interior stencil width of 11, the $p=8$ case has spectral radii that vary a bit over the scheme type, but are not too dissimilar to the lower order operators. Figure~\ref{Fig:D_8-4} indicates convergence rates very near the expected values of $m^{\rm tot}_{L_2}\sim5.5$ and $m^{\rm in}_{L_2}\sim m_{L_\infty}\sim5$.

\emph{The $D^\pm_{9-4}$ Operators } The $p=9$ case shows an improvement of in the spectral radii over the $p=8$ case, practically the same value for all scheme types. Figure~\ref{Fig:D_9-4} shows expected and practically identical convergence properties to the $p=8$ case.

\subsubsection{Dispersion Relation Preserving Operators}

\emph{The $\sD^\pm_{4-2}$ Operators } Requiring an interior stencil width of 9, the DRP $p=4$ scheme shows higher than expected convergence, better even than the $D^\pm_{4-2}$ operator as shown in Figure~\ref{Fig:D_4-2_disp}, at least in the pre-asymptotic regime. This suggests that around this order of accuracy, dispersion error is a significant driver in the overall $L_2$ error.

\emph{The $\sD^\pm_{5-2}$ Operators } The DRP $p=5$ case also exhibits higher than expected convergence, slightly better than that of the $D^\pm_{5-2}$ operators as shown in Figure~\ref{Fig:D_5-2_disp}, at least in the pre-asymptotic regime. This again suggests that around this order of accuracy, dispersion error is a significant driver in the overall $L_2$ error.

\emph{The $\sD^\pm_{6-3}$ Operators } Requiring a interior stencil width of 11, the DRP $p=6$ scheme shows very similar convergence properties to the $D^\pm_{6-3}$ operators as shown in Figure~\ref{Fig:D_6-3_disp}. Notably absent is a higher than expected convergence rate, indicating that the dispersion error is now subdominant.

\emph{The $\sD^\pm_{7-3}$ Operators } The DRP $p=7$ case shows very similar convergence properties to the $D^\pm_{7-3}$ operators as shown in Figure~\ref{Fig:D_7-3_disp}. Here the convergence is again near expected rates, indicating that the dispersion error is again subdominant.

\subsection{Computational Cost}

\begin{table}[tbp]
    \centering
    \begin{tabular}{|c|c|c|c|c|c|c|}\hline\hline
                         & RK Alg. & $h_t/h$ &  $T^{\rm tot} \rm{[mins]}$ & $h^{\rm ext}$ & $T^{\rm ext} \rm{[mins]}$ & $M^{\rm ext} \rm{[GB]}$  \\\hline
        $D^\pm_{2-1}$  & ERK(17,3) & 0.7 & 27  & 0.015   &   $8.5\times 10^{7}$    &  $7.8\times 10^4$   \\
        $D^\pm_{3-1}$  & ERK(17,3) & 2.4 & 9.5  & 0.1   &  $2\times 10^4$    &  $320$  \\
        $D^\pm_{4-2}$  & ERK(17,3) & 1.3 & 24  & 0.28    &  $720$   & 14  \\ 
        $D^\pm_{5-2}$  & ERK(17,3) & 1.7 &  19   &  0.28 &  $600$   & 14 \\ 
        $D^\pm_{6-3}$  & ERK(17,3) & 1.7 &  37  &  0.44   &  170  & 3.4  \\
        $D^\pm_{7-3}$  & ERK(17,3) & 1.6 &  40  &  0.45   &  170  &  3.2  \\
        $D^\pm_{8-4}$  & ERK(18,4) & 1.3 &  $78$   &  $0.65$   & $72$     & $1$  \\
        $D^\pm_{9-4}$  & ERK(18,4) & 1.3 &  $80$ &   $0.65$   &  71  & 1 \\\hline     
        $\sD^\pm_{4-2}$  & ERK(17,3) & 1.1 & 44  &  0.27   &  1500   & 15  \\ 
        $\sD^\pm_{5-2}$  & ERK(17,3) & 0.6 &  80   &  0.29   &  2000   & 12 \\ 
        $\sD^\pm_{6-3}$  & ERK(17,3) & 1.1 &  62  &  0.47   &  210   & 2.7  \\
        $\sD^\pm_{7-3}$  & ERK(17,3) & 0.6 &  113  &  0.49   &  350  &  2.5  \\\hline
    \end{tabular}
    \caption{Comparison of computational cost of all derived operators with an appropriate Runge-Kutta algorithm and observed maximum stable time step size. Here, $T^{\rm tot}$ is the wall clock time for the evolution to ${t=500}$ at the highest tested resolution to complete on an Nvidia RTX 3090 GPU, and $h^{\rm ext}$ is the extrapolated grid spacing required to result in the lowest final value of $E^{\rm tot}_{L_2}$ achieved in this work with the $D^\pm_{9-4}$ operator. Also listed is the extrapolated wall clock time $T^{\rm ext}$ needed to evolve the grid spacing of $h^{\rm ext}$ and $M^{\rm ext}$ is the memory required to hold the state vector of that resolution using 64 bit floating point numbers.}
    \label{Tab:CompCost}
\end{table}

FD based codes, when designed for modern CPUs and GPUs, are usually memory bandwidth limited \cite{omlin2022highperformancexpustencilcomputations,gmd-15-5757-2022}. If this is indeed the case, the computational cost of the implementation of the different operators in this work, keeping the time stepping algorithm fixed, is solely related to how many memory locations must be loaded in order to calculate the right hand side $\partial_tU$, as the time it takes to perform the necessary computations after those memory loads is negligible. In all three of the discussed scheme types, the total number of grid loads needed in 3D is $3p+7$ for even interior order $p$ and $3(p-1)+7$ for odd order, for each grid point in the interior of the domain, neglecting the stencils near the boundaries as they are dwarfed by the interior. This means that the computational cost of the odd order schemes is identical to that of the even order ones under the assumption of a memory bandwidth bottleneck. The fact that the odd order operators tend to have much better spectral radii than their even order counterparts as shown in Figure \ref{Fig:SectralRadii}, and that the convergence properties of the odd operators tend to be better or identical to that of the even ones as shown in Figure~\ref{Fig:total_convergence} and Figure~\ref{Fig:total_disp_convergence}, the choice for reduced computational cost seems to be the odd operators. For asymmetric schemes however, the dispersion error tends to be better for the even non-DRP operators \cite{DualPairingWithDispersionPreserve}. Also, for the DRP operators, the the maximum stable time step was found to be higher for the even operators, at least with the time evolution algorithms used in this study.

To compare the computational cost of operators of different boundary accuracy order $b$, it doesn't usually make sense to keep the time integrator the same, as to achieve the same error tolerance and convergence rate, only an explicit Runge-Kutta (RK) algorithm matching the experimentally observed convergence rate is needed. It is also typical that the spatial discretization error dominates the time discretization error, so sometimes an even lower order RK algorithm is sufficient. As the accuracy order of RK algorithms is increased, more stages are needed, thus increasing the computational cost. To make a useful comparison using the code implementation in this work, all of the operators are compared in Table~\ref{Tab:CompCost} with the fastest observed RK algorithm that retains the same approximate error tolerance and convergence rate as the $\rm{ERK}(20,5)$ algorithm used in the convergence study, along with the largest observed stable time step. Somewhat counterintuitively, high stage count RK methods can lead to faster wall clock times, as the stability region size increases faster than the stage count, allowing larger maximum stable time steps to be taken with fewer stage evaluations per coordinate time. The algorithms used here are the $\rm{ERK}(17,3)$, $\rm{ERK}(18,4)$, and $\rm{ERK}(20,5)$ of \cite{ParsaniOptimizedRK}, named so to indicate the stage count then accuracy order respectively. These RK methods were designed to be optimized for spectral difference operators, and thus may not have optimal stability region shapes for the operators in this work, but nontheless perform much better compared to more traditional RK algorithm choices with smaller stability regions.

The evolution code in this work was written in the Julia programming language \cite{Julia-2017} with the \texttt{ParallelStencil.jl} package \cite{omlin2022highperformancexpustencilcomputations}, allowing it to be run parallelized over CPUs and GPUs. This work utilized one Nvidia RTX 3090 GPU to perform all evolution runs, and a GPU profiling investigation indicated that the code is indeed memory bandwidth limited. Given the final value of the $E^{\rm tot}_{L_2}$ error of the highest resolution tested with the $D^\pm_{9-4}$ operator, each of the lower order operators is compared in Table~\ref{Tab:CompCost} based on the extrapolated grid spacing required $h^{\rm ext}$ to achieve this error tolerance, including the extrapolated wall clock time needed $T^{\rm ext}$ to perform the evolution and the total memory required $M^{\rm ext}$ to hold the state vector of grid spacing $h^{\rm ext}$. Effort can certainly still be expended to optimize the code used in this work for increased performance, but the general trend between operators should largely remain the same.

The results of the computational cost comparison in Table~\ref{Tab:CompCost} unsurprisingly indicate that the higher order operators result in lower wall clock times and lower memory size requirements to perform the evolution for the target error tolerance. It is clear that for a given interior stencil width, DRP operators are not competitive on accuracy or computational cost even though some of them appear to converge faster than expected at least in the pre-asymptotic regime, though they do seem to have slightly more optimal memory requirements. Their use makes more sense however when waves propagate a significant distance on the grid. The DRP operators ensure that the time of flight of these waves over the whole resolvable frequency range is far better approximated than the higher order accuracy operators of the same stencil width, which may result in more accurate solutions as measured at points far away from the sources. A thorough investigation into the cases where DRP schemes are practical in this context is left for future work.

\section{Conclusion and Future Work}\label{Sec:Conclusion}

In this work, new high order embedded boundary finite differencing schemes are derived using the framework of upwind summation by parts operators. These are the highest order embedded boundary finite differencing methods derived to date. The operators are made unique by minimizing the boundary error and are up to $9^{\rm th}$ order accurate in the interior, with the highest boundary closure accuracy of order $4$.

These operators are tested numerically  in 3D using the curvilinear wave equation to demonstrate their convergence properties. The highest achieved convergence rate in this work is demonstrated at order $\sim 5.5$ in the $L_2$ norm of a cubed-sphere 3D multiblock grid setup with an inner embedded spherical boundary.

These operators can be used to evolve hyperbolic problems in first order form with arbitrarily shaped boundaries, while keeping the simplicity of a Cartesian grid structure and the computational efficiency of finite differencing methods. The operators presented in this work can all be reduced to traditional boundary conforming domains by substituting zero for the boundary position parameter, or even facilitate multiple grids that are overset by less than one grid spacing. In particular, these embedded boundary methods can make black hole excision in numerical relativity with finite differencing possible and straight forward to implement without the need for boundary conforming grids while retaining a high order of accuracy. 

Included with this work is some supplementary materials that list the rational polynomial coefficients in the boundary position parameters $\alpha_{l,r}$ for all derived operators as well as a Mathematica notebook for the reader's convenience that can be used to re-derive and symbolically manipulate the solutions.

In the context of the derived operators being applied to the merger of black holes in numerical relativity, the next step left for future work is to investigate these embedded upwind schemes when coupled with the dynamical boundary methods of \cite{GABBARD2024112979}. If dynamical excision boundaries can be achieved on a stationary grid without losing the high convergence order, this could be a compelling alternative to other excision based numerical relativity methods due the absence of the need for a dual-frame evolution scheme. 

The ability to inject BCs via characteristics into the computational domain allows for worldtube excision, a process where a perturbed exact solution is evolved coupled to the numerical domain \cite{Wittek_2023}. The numerical tests performed here suggest this should be possible to achieve on a stationary Cartesian grid, and this is an intended future application.

This work also presented some embedded boundary dispersion relation preserving operators that for a given stencil width, sacrifice accuracy in favor of minimized dispersion error. The use of these operators may significantly reduce the error involved in the long time of flight of waves traveling from the sources generating them, a situation common when generating gravitational waveforms from merger simulations. A thorough investigation into the use of these operators in this context is left for future work.

\begin{acknowledgments}
This work was made possible by valuable discussions with Ken Mattsson, Bernd Br\"ugmann, Sebastiano Bernuzzi, and researchers at the Theoretical Physics Institute at the Friedrich-Schiller University and the Perimeter Institute. Numerical implementation for this project was accomplished thanks to the Julia programming language \cite{Julia-2017} and the packages detailed in \cite{omlin2022highperformancexpustencilcomputations} and \cite{NakamuraTensorial2024}.
\end{acknowledgments}

\bibliography{refs}

\end{document}